\documentclass[a4paper,11pt]{article}
\usepackage{pos}
\usepackage[T1]{fontenc} 
\usepackage{multirow}
\usepackage{tikz}

\usepackage{pdflscape}
\usepackage{booktabs}
\usepackage{verbatim}
\usepackage{amsmath}
\usepackage{graphics}
\usepackage{mathtools}
\usepackage{slashed}
\usepackage{subcaption}

\title{Non-factorisable contributions to $t$-channel single-top production at the LHC and FCC}

\author[a]{Christian Br\o{}nnum-Hansen}
\author*[a]{J\'{e}r\'{e}mie Quarroz}
\author[a,b]{Chiara Signorile-Signorile}
\author[a]{Chen-Yu Wang}

\def\KITA{Institute for Theoretical Particle Physics, KIT, Karlsruhe, Germany}
\def\KITB{Institute for Astroparticle Physics, Hermann-von-Helmholtz-Platz 1, Eggenstein-Leopoldshafen , Germany}

\affiliation[a]{\KITA}
\affiliation[b]{\KITB}

\emailAdd{christian.broennum-hansen@kit.edu}
\emailAdd{jeremie.quarroz@kit.edu}
\emailAdd{chiara.signorile-signorile@kit.edu}
\emailAdd{chen-yu.wang@kit.edu}

\abstract{Single top quark is mainly produced through the $t$-channel W boson exchange $q + b \rightarrow q' + t$ at LHC. This process probes Wtb vertex directly and can be used to measure the CKM matrix element $V_{tb}$ or to constrain the bottom quark PDF. The non-factorisable contributions are the last missing piece of the NNLO corrections. In these proceedings, we discuss in a first part the ostensible importance of such corrections and the obtention of the different relevant amplitudes. In the second part, recently published results at the energy of the LHC are compared to new results for proton-proton collisions at $100\, {\rm TeV}$, the energy of the FCC.}

\FullConference{%
  Loops and Legs in Quantum Field Theory - LL2022,\\
  25-30 April, 2022\\
  Ettal, Germany
}

\def\@preprint{\@empty}
\newcommand\preprint[1]{\gdef\@preprint{\hfill #1}}
\preprint{TTP22-050, P3H-22-080}

\begin{document}
\noindent\@preprint\par
\maketitle

\section{Introduction}

The physics of the top quark is unique. Indeed, the lifetime of the top quark, the heaviest particle of the Standard Model, is so short that it does not hadronise. As a consequence, non-perturbative effects do not need to be considered to get accurate predictions for processes in which top quarks are produced at colliders. In addition, since the top-quark coupling to the Higgs boson is close to one, its study may provide important insights into the electroweak symmetry breaking.

At the LHC, top quarks are mostly produced in $t\bar{t}$ pairs.
Nevertheless, single-top production also happens frequently. Indeed, the total cross-section for single-top production is about four times smaller than the one for $t\bar{t}$ pair production.
The NLO QCD and NLO electroweak corrections to single-top production with decay of the top quark are known since more than ten years now~\cite{Bordes:1994ki,Campbell:2004ch,Cao:2004ky,Cao:2005pq,Harris:2002md,Schwienhorst:2010je}. More recently the NNLO QCD corrections have been computed in the factorisable approximation~\cite{Brucherseifer:2014ama,Berger:2016oht,Berger:2017zof,Campbell:2020fhf}. The mixed QCD-electroweak corrections and soft-gluon resummation have also been studied~\cite{Cao:2018ntd, Frederix:2019ubd}.

The single-top quark production is interesting as it allows a direct determination of the CKM matrix element $V_{bt}$~\cite{Rindani:2011pk, Cao:2015doa,Gonzalez-Sprinberg:2015dea,CDF:2015gsg,CMS:2014mgj}. This leads to an indirect determination of the decay rate of the top quark as it will mostly decay into a bottom quark by emitting a W boson. In addition, it offers a way to constrain the bottom-quark PDF.

At hadron collider, single top quarks can be produced in three different ways with rather different rates. Indeed, $70\%$ of the single-top quarks at the LHC originate from the so-called $t$-channel production mode. It will be the subject of our study.

In these proceedings, we discuss the non-factorisable corrections to the $t$-channel single-top quark production. In the first part, we discuss the ostensible importance of such corrections and the obtention of the different relevant amplitudes. In the second part, recently published results at the energy of the LHC are compared to new results for proton-proton collisions at $100\, {\rm TeV}$, the energy of the FCC~\cite{FCC:2018vvp,Mangano:2017tke}.

\section{Non-factorisable corrections}

\begin{figure}[ht]
    \centering
    \begin{subfigure}[ht]{0.42\linewidth}
        \centering
        \includegraphics[height=2.9cm]{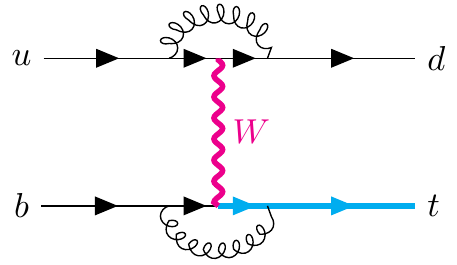}
        \subcaption{Factorisable diagram contributing to the double-virtual amplitude.}
    \end{subfigure}
    \hspace{1cm}
    \begin{subfigure}[ht]{0.42\linewidth}
        \centering
        \includegraphics[height=2.6cm]{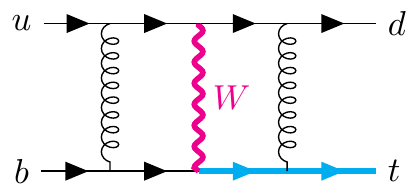}
        \subcaption{Non-factorisable diagram contributing to the double-virtual amplitude.}
    \end{subfigure}
    \caption{In the factorisable approximation, diagrams made of a cross-talk between the two fermion lines are neglected since they are suppressed by a factor $N_c^2-1 = 8$ at NNLO compared to diagrams where the corrections are confined to the same quark line.}
    \label{fig:ex}
\end{figure}

The Born amplitude for the $t$-channel single-top quark production is made of two fermion lines connected by a $W$ boson. Non-factorisable corrections are described by diagrams where gluons are exchanged between these two fermions line, see Fig.~\ref{fig:ex}.
On the left, the gluons are emitted and absorbed by the same quark line and therefore the diagram is part of the factorisable corrections.
On the right, a diagram contributing to the non-factorisable two-loop amplitude is shown.
The non-factorisable corrections are usually neglected because they are colour-suppressed~\cite{Brucherseifer:2014ama,Campbell:2020fhf}. It is easy to compute the color factor of these two diagrams once they are projected on the Born amplitude. The resulting color factors are $\frac1{4}(N_c^2-1)^2$ for the factorisable diagram and $\frac1{4}(N_c^2-1)$ for the non-factorisable one respectively. Therefore, the non-factorisable corrections are suppressed by a factor $N_c^2-1 = 8$.
In addition, one cannot rely on an estimate of these corrections at NLO since the different interferences vanish at this order because of colour conservation.

Nevertheless, there are several reasons to believe that the non-factorisable should not be ignored. First of all, the NNLO QCD corrections to the $t$-channel single-top production in the factorisable approximation have been computed recently and appear to be small, namely $\mathcal{O}(1\%)$~\cite{Campbell:2020fhf}. In addition, there could also be a $\pi^2$ enhancement due to the Glauber or Coulomb phase. This is a virtual effect which is linked to the infinite perturbative interaction range between the two quark lines exchanging gluons. In principle, it does not require a scattering to occur. This effect has been studied in the context of two electrons exchanging an arbitrary number of photons in the limit where the centre-of-mass energy is large compared to the momentum transfer~\cite{Chang:1969by}. We note that the presence of the $\pi^2$ enhancement has been explicitly shown for the non-factorisable NNLO QCD corrections to the vector boson fusion in the eikonal approximation~\cite{Liu:2019tuy}. If this enhancement factor $\pi^2 \sim 10$ exist for the single-top production, it could clearly compensate for the colour suppression.

There is another consequence of this effect; namely that the role of the virtual corrections is expected to be dominant in non-factorisable corrections. This can be shown by expanding the integrated cross section in the transverse momentum of the top quark $p_\perp^t$
\begin{equation}
    \sigma = \sigma_0 + \frac{p_\perp^t}{\sqrt{s}}\: \sigma_1 + \mathcal{O}\left(\left(p_\perp^t/\sqrt{s}\right)^2\right)
    \,.
    \label{eq:exp_sigma}
\end{equation}
This expansion is valid for the $t$-channel single-top production as the typical value of the transverse momentum of the top-quark is $p_\perp^t \sim 40\: {\rm GeV}$ and $\sqrt{s}\sim 300\: {\rm GeV} $. As the $\pi^2$ enhancement occurs when $p_\perp^t\rightarrow 0$, it will affect the term $\sigma_0$ in Eq.~\eqref{eq:exp_sigma}. However, in this limit, the real emission vanishes up to corrections due to the mass of the top quark. Indeed, to contribute to $\sigma_0$, the emission has to be collinear or soft. Since non-factorisable amplitude are free of collinear divergences, we only need to consider emission of a soft gluon. In this limit, any amplitude factorises. For instance, the five-point amplitude can be written as
\begin{align}
    \begin{split}
    \lim_{E_5 \rightarrow 0} \mathcal{A}^{(0)}(1_q,2_b,3_q',4_t,5_g) =& -g_s^2 C_F (\frac{p_1^\mu}{k\cdot p_1} - \frac{p_3^\mu}{p_5\cdot p_3} + \frac{p_2^\mu}{p_5\cdot p_2} - \frac{p_4^\mu}{p_5\cdot p_4})\epsilon^*_\mu(p_5)\\
    &\times\mathcal{A}^{(0)}(1_q,2_b,3_q',4_t)
    \,,
    \end{split}
\end{align}
where $\mathcal{A}^{(0)}(1_q,2_b,3_q',4_t)$ is the Born amplitude and $\epsilon^*_\mu(p_5)$ is the polarisation vector for the emitted gluon $5_g$. We see that, in the limit $p^t_\perp \rightarrow 0$ and for massless momenta, the soft function vanishes as $p_1^\mu = p_3^\mu$ and $p_2^\mu = p_4^\mu$. In the case of single-top production, $p_4^\mu$ is massive and this argument holds only when the centre-of-mass energy is much larger than the top mass.

\section{Real-virtual amplitude}
The calculation of the real-virtual amplitude turns out to be non-trivial due to the presence of multiple mass scales. We generate the 24 diagrams with QGRAF and we treat them with FORM. We work in the 't Hooft-Veltman scheme where the external momenta live in four dimensions and the internal ones live in $d=4-2\epsilon$. The $\epsilon -$dependence of the different spinor structures can be extracted by splitting the gamma matrices into four-dimensional and $\epsilon$-dimensional parts
\begin{equation}
    \gamma^\mu = \gamma^{\bar{\mu}} + \gamma^{\tilde{\mu}}
    \,,
\end{equation}
where
\begin{equation}
    \{\gamma^{\bar{\mu}},\gamma_5\} = 0 \qquad \qquad [\gamma^{\bar{\mu}},\gamma_5] = 0
    \,.
\end{equation}
It is now clear that one needs at least two gamma matrices living in $-2\epsilon$ space to produce an $\epsilon$ term. For instance, the following spinor structure
\begin{equation}
    \bar{u}_t(p_4) \gamma^\mu \gamma^\nu u_b(p_2) = \bar{u}_t(p_4) \gamma^{\bar{\mu}} \gamma^{\bar{\nu}} u_b(p_2) + g^{\tilde{\mu}\tilde{\nu}}\bar{u}_t(p_4)u_b(p_2)
    \,,
\end{equation}
can be expressed as pure four dimensions spinor chains. We projected the two matrices $\gamma^{\tilde{\mu}}\gamma^{\tilde{\nu}}$ on the metric tensor $g^{\tilde{\mu}\tilde{\nu}}$ as it is the only object that can be build from two indices living in $-2\epsilon$ dimension provided that all other momenta are purely four-dimensional.

The treatment of $\gamma_5$ within 't Hooft-Veltman scheme is generally difficult. However, the case of the single-top production is convenient. Indeed, the left-handed projector from the quark-quark-W boson vertex can be moved to the external states from the beginning. Then, we decide to remove these projectors and to work with helicity amplitudes in $d$ dimension. Finally, we select only the left-handed external state.

The next step is to decompose the massive momentum $p_4$ into a reference vector $r$ and a new massless momentum $4^\flat$
\begin{equation}
    P_L u_t(p_4) = |4^\flat] + \frac{m_t}{\langle 4^\flat r \rangle} |r\rangle \qquad \qquad P_R u_t(p_4) = | 4^\flat \rangle + \frac{m_t}{[4^\flat r]} | r]
    \,,
\end{equation}
where $P_L=\frac{1-\gamma_5}{2}$ stands for the left-handed projector and $P_R=\frac{1+\gamma_5}{2}$ the right-handed one. The reference vector $r$ is an arbitrary massless momentum. It can be choosen conveniently to simplify the expressions.
After simplification, we are left with two spinor structures per helicity configuration. There are four of them as both the emitted gluon and the top quark can be left- or right-handed.

On the other hand, one needs to treat the form factors. The challenge comes from the reduction of five-point rank-three integrals. As an example, we consider a rank $r\le 3$ tensor pentagon integral
\begin{equation}
                \int \frac{d^dk}{(2\pi)^d}\frac{k^{\mu_1}\dots k^{\mu_r}}{\prod_{i=1}^5 \left[(k+q_i)^2-m_i^2\right] } 
                \,,
\end{equation}
where all momenta are considered to be incoming, so that $\sum_{i=1}^5 p_i = 0$, and the flowing momenta are defined as $q_i = \sum_{j=1}^i p_j $. We introduce the \emph{van Neerven-Vermaseren} basis $\{v_i\}_{i=1\dots 4}$ which spans the physical momentum space. We refer the reader to the following reference for a detailed discussion~\cite{Ellis:2011cr}. The basis vectors are defined to fulfill the following relation
\begin{equation}
    v_i\cdot p_j = \delta_{ij}
    \label{eq:vNV_prop}
    \,.
\end{equation}
We emphasise that, since the five incoming momenta are not independent, the dot product of the vector $v_i$ to the left-over momentum is fully-defined by the four others. The loop momentum can be expressed in this new basis as
\begin{equation}
    k^\mu = \sum_{i=1}^4 (k\cdot p_i)v_i^\mu + k^{\tilde{\mu}}
            \,,
\end{equation}
where $k^{\tilde{\mu}}$ stands for the $-2\epsilon$-dimension part of the loop momentum $k^\mu$. Since five-point rank-three integrals do not contain rational terms, the former can be dropped.
The factor $k\cdot p_i$ can be expressed as
\begin{equation}
            k\cdot p_i =
            \frac1{2}\left[(k+q_i)^2-m_i^2\right] - \frac1{2}\left[(k+q_{i-1})^2-m_{i-1}^2\right]
                + \frac1{2}\left[m_i^2 - m_{i-1}^2 -p_i^2- 2p_i\cdot q_{i-1}\right]
                \,.
                \label{eq:kpi}
\end{equation}
The first two terms in Eq.~\eqref{eq:kpi} lead to two rank $r-1$ boxes which we can treat with a standard Passarino-Veltman reduction. The last term corresponds to a pentagon of rank $r-1$; to deal with it, one can repeat this procedure. Once we reached the point of scalar pentagon, they can be expressed as a combination of five scalar box integrals, one for each pinched propagator of the pentagon integrals up to $\mathcal{O}(\epsilon)$ terms~\cite{Bern:1993kr}.

The real-virtual amplitude is now expressed in terms of 109 scalar box, triangle and bubble integrals. To improve the numerical stability of the amplitude, we want to reduce the size of the integral coefficients. The most complicated ones come with the box integrals. We decided to switch to a basis of finite box integrals in order to set the dimensional regulator present in their coefficient to zero. As an example, we consider the following box integral
\begin{equation}
                I_{4,1} = \int \frac{\mathrm{d}^dk}{(2\pi)^d}\, \frac{1}{ k^2   (k - p_1)^2 (k - p_1- p_2)^2   (k - p_1- p_2 + p_5 )^2}\;.
                \label{eq:I41}
\end{equation}
This integral is divergent as any of the propagator goes on-shell. It can be regulated through a numerator insertion which vanishes in these same limits
\begin{align}
    \mathrm{tr} \left( (-\slashed{p}_1) (\slashed{k} - \slashed{p}_1) (\slashed{k} - \slashed{p}_1 - \slashed{p}_2) (\slashed{p}_5) \right) =\, &-s_{12}\, (s_{12} + s_{15} - s_{34}) + (s_{12} + s_{15} - s_{34})\,  k^2 \nonumber \\                                                                                     &- (s_{12} - s_{34})\,  (k - p_1)^2 +(s_{12} + s_{15})\,  (k - p_1 - p_2)^2  \nonumber \\
                & - s_{12}\,  (k - p_1 - p_2 + p_5)^2 \,.
                \label{eq:trace}
\end{align}
After the trace is calculated, we observe that four terms on the right-hand side of Eq.~\eqref{eq:trace} correspond to the four different propagator of the considered box integrals Eq.~\eqref{eq:I41}. These terms will lead to triangle integrals. The last term is independent of $k$. This factor multiplies the initial divergent box integral. Using this relation, we can therefore express the divergent box integral as a sum of one finite box integral that involves the trace that appear on the left-hand side of Eq.~\eqref{eq:trace} and four divergent triangle integrals. This redefinition can be done for any of the divergent boxe integrals present in the amplitude.

As all the divergences are now in the triangle integrals and since the pole structure for non-factorisable corrections at NNLO is very simple, we observe that the coefficients of the triangle integrals either become independent of the dimensional regulator $\epsilon$ or simply vanish.

\section{Double-virtual amplitude}
The double-virtual amplitude is obtained numerically through the auxiliary mass flow method. We refer the reader to Ref.~\cite{Bronnum-Hansen:2021pqc} for a detailed explanation. The analytic reduction of the two-loop amplitude keeping the full dependence on the two Mandelstam variables $s$ and $t$ and the two masses $m_W$ and $m_t$ is possible within 4 days on 20 cores. We end up with 428 masters integrals split into 18 families. The evaluation of the master integrals through the auxiliary mass flow method can be performed to any desired accuracy. We were able to evaluate the 428 master integrals at a typical phase space point to 20 digits accuracy within 30 minutes on a single core.
 \begin{table}[t]
     \begin{center}
         \begin{tabular}{|c|c|c|}
 	\hline
 	& $\epsilon^{-2}$ & $\epsilon^{-1}$ \\
 	\hline \hline
             $\langle\mathcal{A}^{(0)}|\mathcal{A}_{\rm nf}^{(2)}\rangle $ & \scriptsize{$-229.094040865466{\color{red}0} -8.978163333241{\color{red}640} i$} & \scriptsize{$-301.18029889447{\color{red}64} -264.17735965295{\color{red}05} i$}  \\
 	IR poles & \scriptsize{$-229.0940408654665 -8.978163333241973 i$} & \scriptsize{$-301.1802988944791 -264.1773596529535 i$} \\
 	\hline
         \end{tabular}
     \end{center}
     \caption{Comparison of the poles of the two-loop amplitude with the ones predicited using Catani's operator.}
     \label{tab:poleVV}
 \end{table}
 In table Tab.\ref{tab:poleVV}, we present the poles obtained for the two-loop amplitude compared to the one predicted by Catani's operator~\cite{Catani:1998bh}. We observe that we have about 15 digits match at $\epsilon^{-2}$ and 14 digits match at $\epsilon^{-1}$. We expect therefore that the finite part is correct to about 13 digits, as we lose one digit per $\epsilon$ order.

 To evaluate the cross section, we prepare a Vegas grid starting with the Born squared cross section. We extract then 10 sets of $10^4$ points from this grid and we evaluate the amplitude for each of these points. Since we have 10 independent sets, we can use the spread of the NNLO double-virtual correction to estimate the accuracy of our result, namely $\mathcal{O}(2\%)$.

\section{Results at the LHC}

We consider proton-proton collisions at 13 TeV. We use the PDF set \texttt{CT14}. The leading-order cross section is computed with then leading-order PDFs whereas the NNLO cross section is computed with NNLO PDFs.
We present the integrated cross section at a fixed factorisation scale $\mu_F = m_t$
\begin{equation}
                \frac{\sigma_{pp\rightarrow X + t}}{1\,{\rm pb}} = 117.96 + 0.26\left(\frac{\alpha_s(\mu_R)}{0.108}\right)^2
                \,.
\end{equation}
At $\mu_R = m_t$, the non-factorisable corrections to the leading-order cross section amount to $0.2\%$. The non-factorisable cross section displays a trivial dependence on the renormalisation scale. As the non-factorisable corrections appear for the first time at NNLO, they are independent of LO, NLO and NNLO \emph{factorisable} contributions. As a consequence, at this order, there is no indication of a good scale choice. Analogously to deep inelastic scattering, the renormalisation scale may be chosen to be the typical transverse momentum of the top quark, $\mu_R = 40\, {\rm GeV}$. In this case, the non-factorisable corrections become close to $0.35\%$. In comparison, the NNLO factorisable corrections to the NLO cross section are about $0.7\%$~\cite{Campbell:2020fhf}.

\begin{figure}[t]
    \centering
      \begin{subfigure}{0.495\textwidth}
    \centering
    \includegraphics[width=\textwidth]{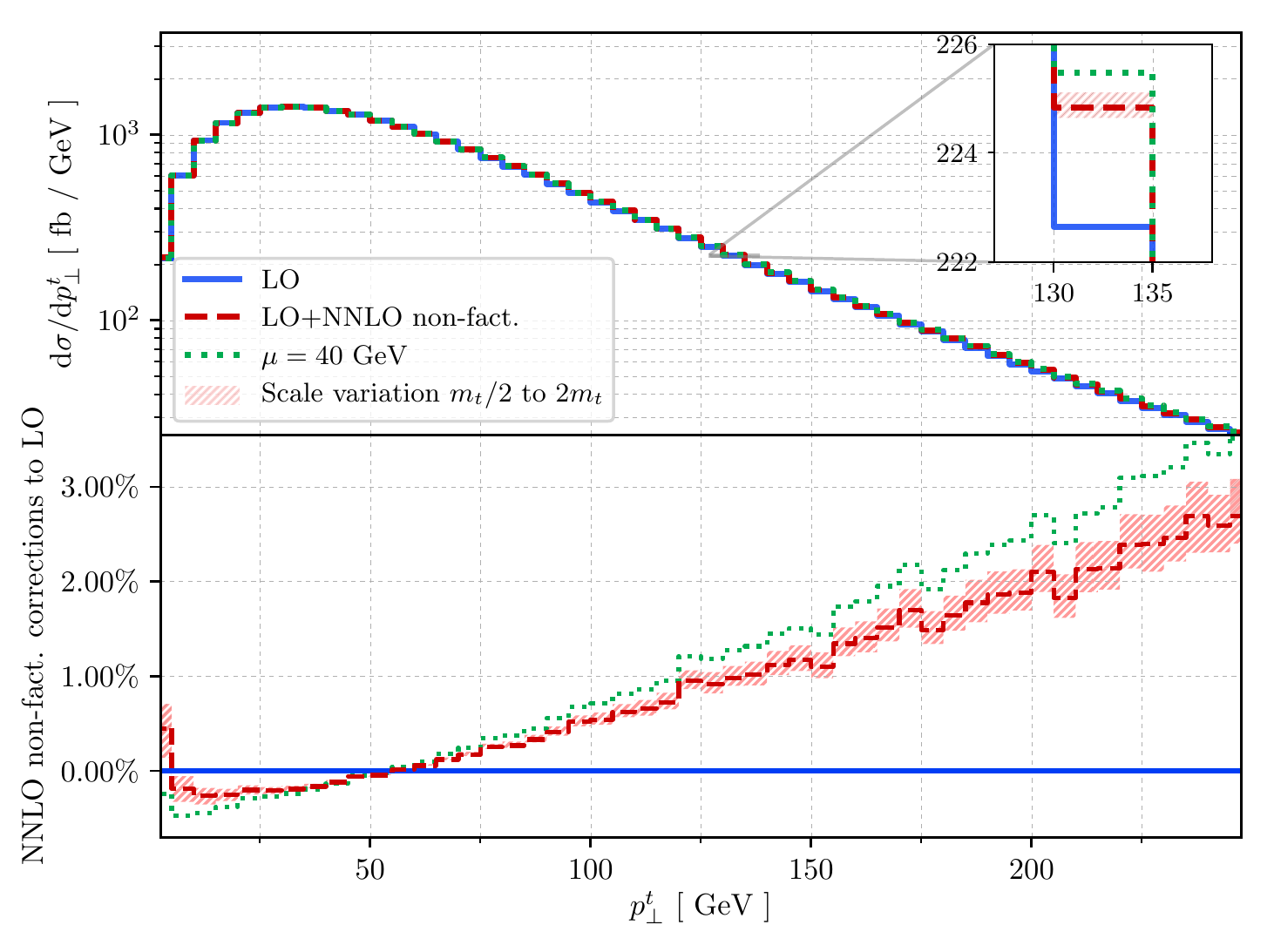}
    \caption{13 TeV}
    \label{fig:pttop-13}
  \end{subfigure}
  \begin{subfigure}{0.495\textwidth}
    \centering
    \includegraphics[width=\textwidth]{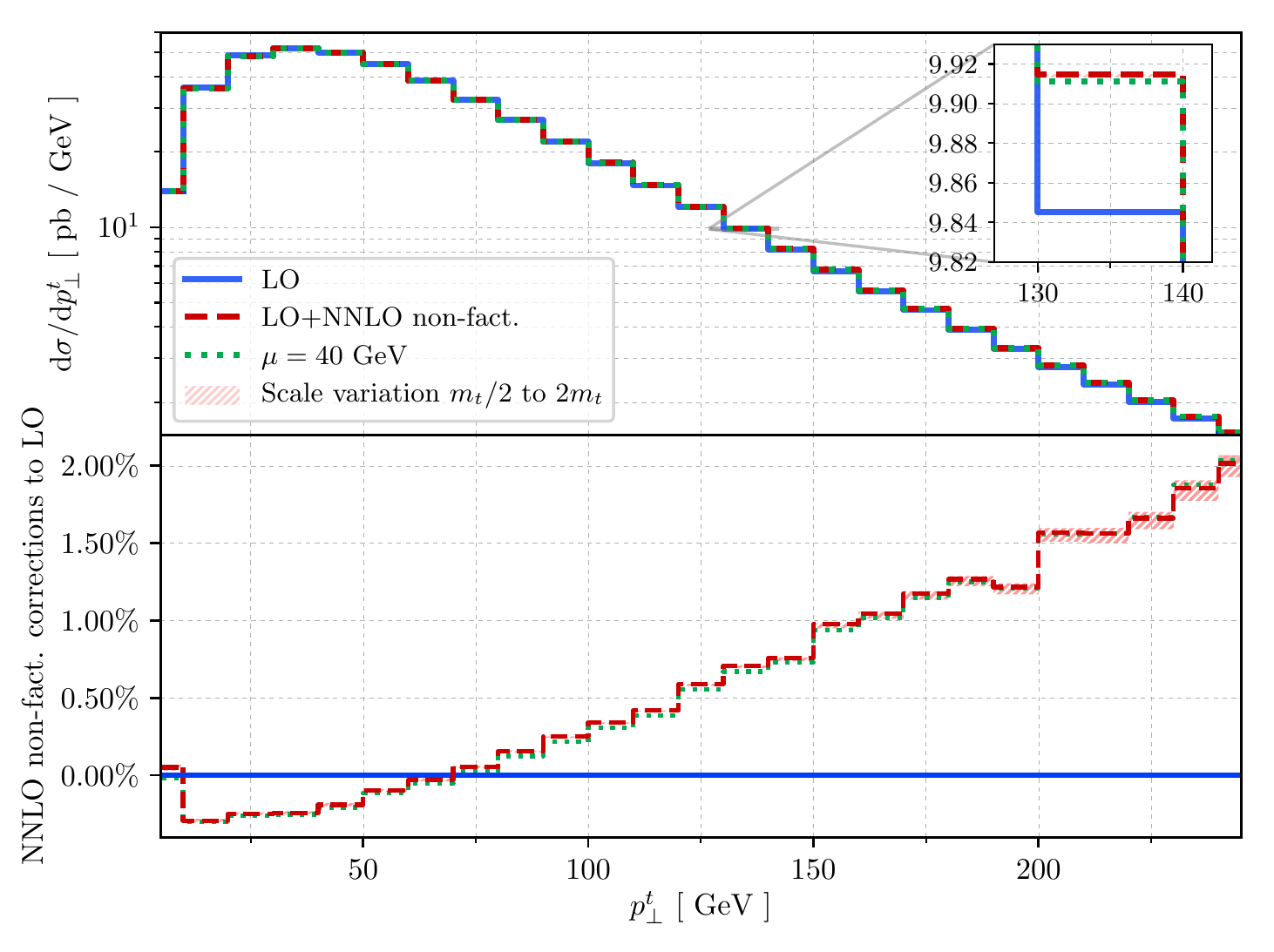}
    \caption{100 TeV}
    \label{fig:pttop-100}
  \end{subfigure}
\centering
\caption{The top quark transverse momentum distribution. The blue solid line represents the Born cross section. The red dashed line corresponds to the non-factorisable corrections at $\mu \equiv \mu_F = \mu_R = m_t$. The variation of $\mu$ by a factor 2 is depicted by the red region around the red dashed line. We also present with a green dotted line at the scale $\mu = 40\; {\rm GeV}$.}
\label{fig:pttop}
\end{figure}

The impact of the non-factorisable corrections on the top-quark transverse momentum distribution is presented on the left pane in Fig.~\ref{fig:pttop}. The blue solid line represents the Born cross section. The red dashed line corresponds to the non-factorisable corrections at $\mu \equiv \mu_F = \mu_R = m_t$. The variation of $\mu$ by a factor 2 is depicted by the red region around the red dashed line. A green dotted line shows NNLO corrections computed with the scale $\mu = 40\; {\rm GeV}$.
We observe that the non-factorisable corrections exhibit a significant dependendence on the transverse momentum of the top quark.
Another interesting feature is that the non-factorisable corrections vanish around at $50 \, {\rm GeV } $ for any scale value. On the other hand, the factorisable corrections vanish at $30 \; {\rm GeV}$~\cite{Campbell:2020fhf}. The consequence is that there is a part of the phase space at low $p_t^{\perp}$, close to the peak of the $p_t^{\perp}$ distribution, where the non-factorisable corrections are \emph{dominant} compared to factorisable corrections.  

\section{Non-factorisable corrections at the FCC}

In this section, we present results for proton-proton collision at $100\: {\rm TeV}$. We use the same PDF set as in the previous section. The integrated cross-section reads
\begin{equation}
                \frac{\sigma_{pp\rightarrow X + t}}{1\,{\rm pb}} = 2367.0 + 3.8\left(\frac{\alpha_s(\mu_R)}{0.108}\right)^2
                \,.
\end{equation}
At $100 \, {\rm TeV}$, we observe that at $\mu_R = m_t$, the NNLO corrections amount to $0.16\%$. At $\mu_R=40 \, {\rm GeV}$, these corrections become $0.25\%$.
\begin{table}
    \centering
\begin{tabular}{@{} ll|ll|ll @{}}    \toprule
     & & \multicolumn{2}{c}{$\mu_R = m_t $}  & \multicolumn{2}{|c}{$\mu_R = 40\: {\rm GeV}$} \\ 
    $p_{\perp}^{t,\text{cut}}$ &  $\sigma_{\text{LO}}$ (pb) & $\sigma^{{\rm nf}}_{\text{NNLO}}$ (pb) & $\delta_{\text{NNLO}} \: [\%]$ & $\sigma^{{\rm nf}}_{\text{NNLO}}$ (pb) & $\delta_{\text{NNLO}}\: [\%]$\\ \midrule 
 0 GeV & $2367.02$  & $3.79_{\;\;\:0.84}^{-0.63}$ & $0.16_{\;\;\:0.04}^{-0.03} $ & $5.95$ & $0.25$ \\ 
 20 GeV & $2317.03$  & $3.89_{\;\;\:0.86}^{-0.64}$ & $0.17_{\;\;\:0.04}^{-0.03} $ & $6.11$ & $0.26$ \\ 
 40 GeV & $2216.61$  & $4.14_{\;\;\:0.92}^{-0.69}$ & $0.19_{\;\;\:0.04}^{-0.03} $ & $6.50$ & $0.29$  \\ 
 60 GeV & $2121.88$  & $4.28_{\;\;\:0.95}^{-0.71}$ & $0.20_{\;\;\:0.04}^{-0.03} $ & $6.71$ & $0.32$ \\  \bottomrule 

\hline
\end{tabular}
\caption{Born and NNLO non-factorisable cross-sections computed with a cut on the transverse momentum of the top quark at fixed factorisation scale $\mu_F=m_t$ at $100\, {\rm TeV}$. The first column corresponds to the value of the cut. The second column gives the leading-order cross section. The third and the fourth column present the NNLO non-factorisable corrections at scale $\mu_R=m_t$ with its variation by a factor 2. The two last column describes the same quantities at renormalisation scale $\mu_R=40\, {\rm GeV}$.}
\label{tab:pttop}
\end{table}


In the right pane of Fig.~\ref{fig:pttop}, the distribution of the top-quark transverse momentum is plotted. We emphasise that we change unit from ${\rm fb}$ at $13 \, {\rm TeV}$ to ${\rm pb}$ on this figure at $100 \, {\rm TeV}$. The shape of the distribution is similar. The peak of the distribution is still centred around $p_t^\perp = 40\, {\rm GeV}$. The NNLO non-factorisable corrections change sign around $70\, {\rm GeV}$ whereas at $13 \, {\rm TeV}$, it was at $50\, {\rm GeV}$.

In Tab.~\ref{tab:pttop}, we present the leading-order cross section and the NNLO non-factorisable cross section subject to different cuts on the transverse momentum of the top quark. The factorisation scale is fixed to $\mu_F=m_t$. In the first two columns, we present the leading-order cross section at four different cut values. The third and the fourth column corresponds to $\mu_R=m_t$. We vary the renormalisation scale by a factor 2. The last two columns corresponds to $\mu_R=40 \, {\rm GeV}$.
\begin{figure}[t]
    \centering
      \begin{subfigure}{0.495\textwidth}
    \centering
    \includegraphics[width=\textwidth]{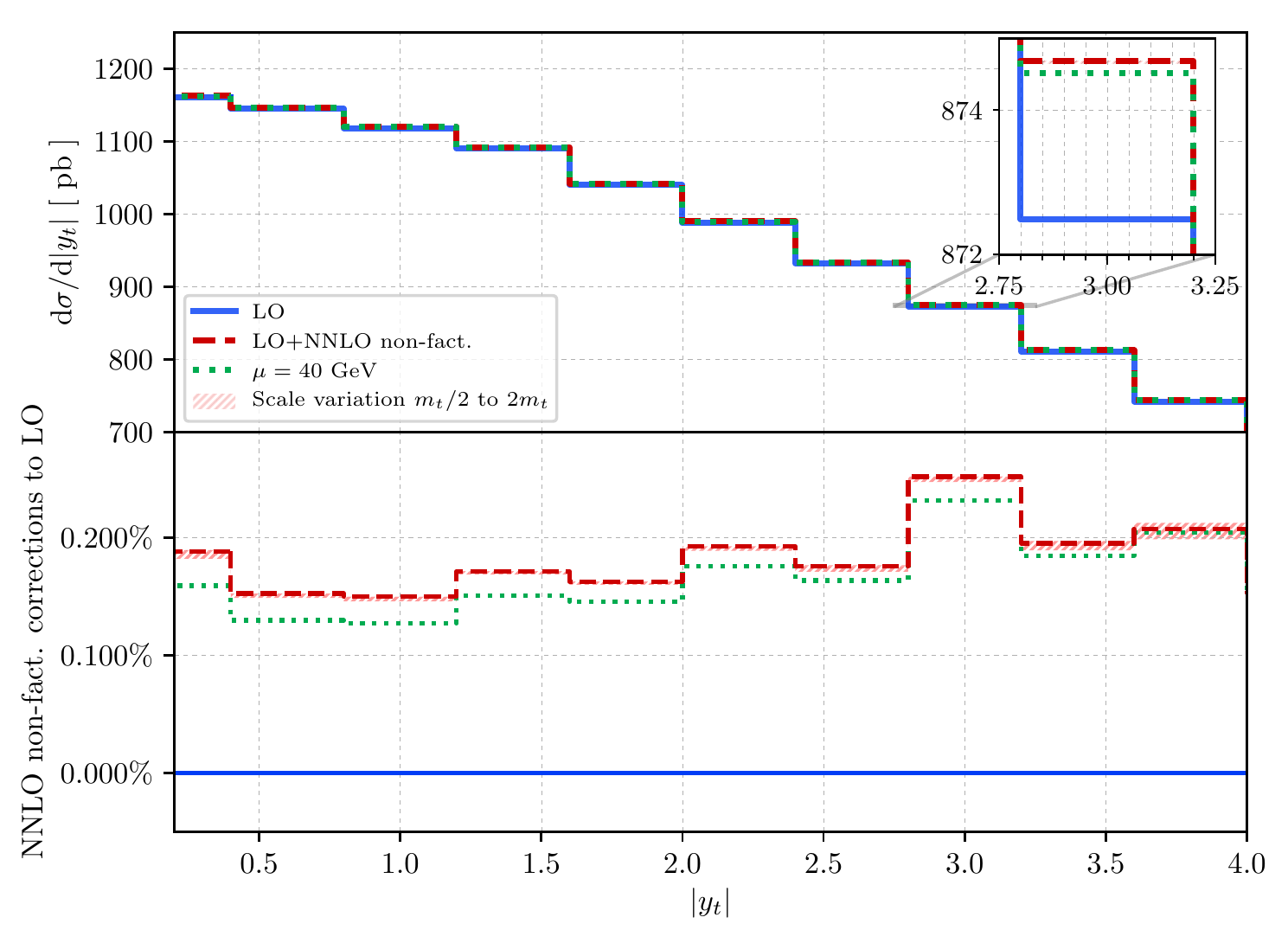}
    \caption{Distribution of the top-quark rapidity.}
    \label{fig:ytop}
  \end{subfigure}
  \begin{subfigure}{0.495\textwidth}
    \centering
    \includegraphics[width=\textwidth]{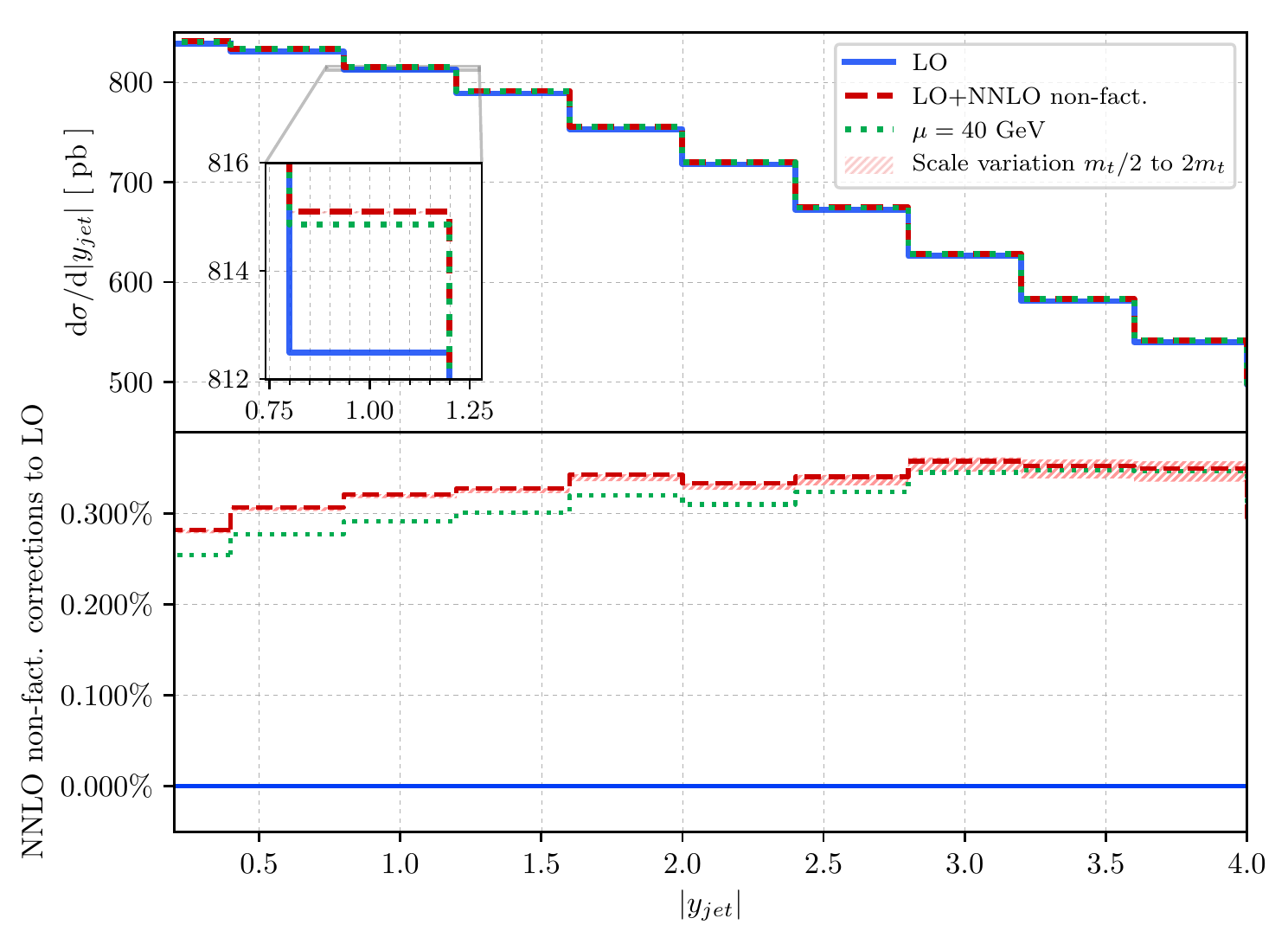}
    \caption{Distribution of the leading-jet rapidity.}
    \label{fig:yjet}
  \end{subfigure}
  \caption{Distribution of the rapidity of the top quark (left pane) and of the leading jet (right pane) at 100 TeV. The red dashed line corresponds to the scale $\mu \equiv \mu_F = \mu_R = m_t$. The green dotted line to $\mu=40 \, {\rm GeV}$.}
  \label{fig:ytopandyjet}
\end{figure}

In Fig.~\ref{fig:ytopandyjet}, we present the rapidity of the top quark on the left pane and the rapidity of the leading jet on the right pane. The corrections are nearly constant around $0.2\%$. On the right pane, one can read the distribution of the leading-jet rapidity.  The corrections to this observable turn out to be also almost flat and amount to $0.35\%$. The behaviour of these two rapidity distributions is different from the ones at $13 \, {\rm TeV}$. Indeed, at this centre-of-mass energy, the corrections are drastically decreasing around $y=2$ and even negative for rapidities above $y=3.5$~\cite{Bronnum-Hansen:2022tmr}.
\begin{figure}[th]
    \centering
      \begin{subfigure}{0.495\textwidth}
    \centering
    \includegraphics[width=\textwidth]{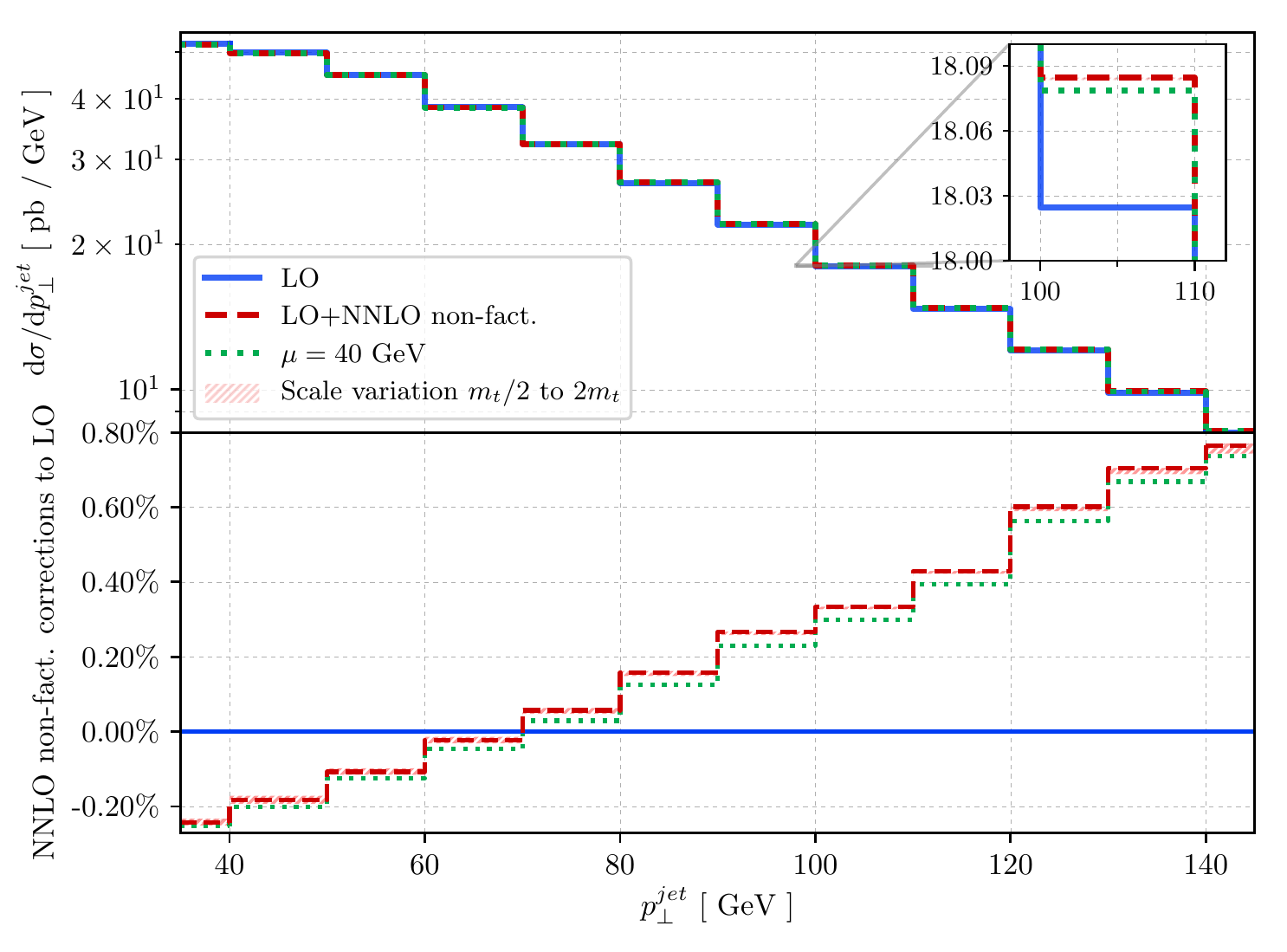}
    \caption{Leading-jet transverse momentum.}
    \label{fig:ptjet}
  \end{subfigure}
  \begin{subfigure}{0.495\textwidth}
    \centering
    \includegraphics[width=\textwidth]{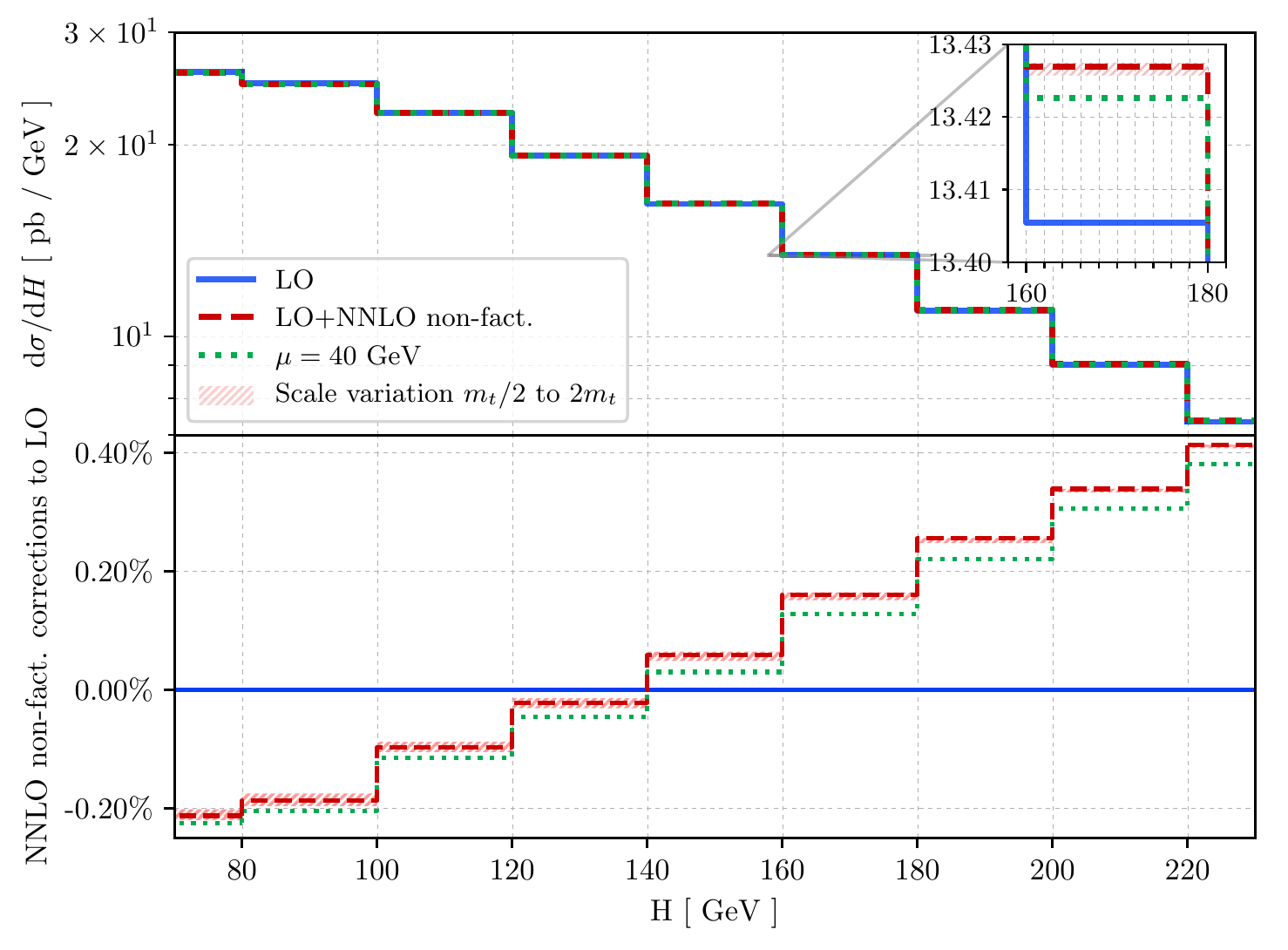}
    \caption{Sum of transverse momenta $H$.}
    \label{fig:H}
  \end{subfigure}
  \caption{Distribution of the transverse momentum of the leading jet (left pane) and of the sum of the transverse momenta defined in Eq.~\eqref{eq:Hdef} (right pane) at 100 TeV. The red dashed line corresponds to the scale $\mu \equiv \mu_F = \mu_R = m_t$. The green dotted line to $\mu=40 \, {\rm GeV}$.}
  \label{fig:ptjetandH}
\end{figure}

In Fig.~\ref{fig:ptjetandH}, we present the distribution of the leading jet transverse momentum on the left pane and the sum of the transverse momenta
\begin{equation}
    H = \sum_{i=1}^{n_{\text{jet}}} p_\perp^{\text{jet}, i}
    \label{eq:Hdef}
    \,,
\end{equation}
on the right pane. The shape of the NNLO non-factorisable corrections on these two observables is similar to the ones at $13\, {\rm TeV}$. It is interesting to note that the hierarchy between the different scales is inverted compared to $13 \, {\rm TeV}$ results. Indeed, in Fig.~\ref{fig:ptjetandH}, the corrections at $\mu=40 \, {\rm GeV}$ represented by a green dotted line are smaller than the ones at $\mu = m_t/2$, $\mu=m_t$ and $\mu=2m_t$ in red dashed line.

\section{Conclusions}
We report on the computation of \emph{non-factorisable} corrections to the $t$-channel single-top production.
The double-virtual amplitude has been numerically evaluated using the auxiliary mass flow method. This procedure is sufficiently robust to produce phenomenologically relevant results.
Due to multiple mass scales appearing in the one-loop five-point amplitude, its reduction to master integrals and its stable and efficient numerical evaluation turn out to be non-trivial.
Since the NNLO factorisable corrections to the NLO cross section are small, the non-factorisable corrections are found to be quite comparable.
We provide new results for non-factorisable corrections for proton-proton collisions at $100\, {\rm TeV}$. The integrated cross-section is changed by few permilles and the corrections to the different distributions have the tendency to be much flatter than at $13\, {\rm TeV}$. It is hard to estimate how the factorisable corrections will be affected by such an increase in the beam energy through an heuristic argument. Such results would be desirable to understand the enhancement of the non-factorisable corrections when the centre-of-mass energy becomes large.

\bibliographystyle{JHEP}
\bibliography{references}


\end{document}